%% file: main.tex
\documentclass[conference]{IEEEtran}
\IEEEoverridecommandlockouts
\usepackage{cite}
\usepackage{amsmath,amssymb,amsfonts}
\usepackage{algorithmic}
\usepackage{graphicx}
\usepackage{textcomp}
\usepackage{xcolor}
\usepackage{booktabs} 
\usepackage{enumitem}

\usepackage{etoolbox}
\usepackage{fancyhdr}
\usepackage{soul,xcolor}
\usepackage{pgfplots}
\usepackage{subfloat}
\usepackage{subfig}
\usepackage{listings}
\usepackage{cleveref}
\usepackage{xspace}
\usepackage{makecell}
\usepackage{multirow}
\usepackage{url}

\newcommand{\arch}{\textsc{TCX}\xspace}
\newcommand{\vm}{\textsc{SC-VM}\xspace}
\newcommand{\vms}{\textsc{SC-VMs}\xspace}
\newcommand{\vendorvm}{Root \textsc{VM}\xspace}

\definecolor{grey1}{HTML}{d9d9d9}
\definecolor{grey2}{HTML}{a6a6a6}
\definecolor{grey3}{HTML}{6c6c6c}
\definecolor{grey4}{HTML}{404040}
\definecolor{black}{HTML}{000000}

\newcommand*\circled[1]{\tikz[baseline=(char.base)]{
            \node[shape=circle,draw,inner sep=.2pt] (char) {\small{#1}};}}

\colorlet{punct}{red!60!black}
\definecolor{background}{HTML}{EEEEEE}
\definecolor{delim}{RGB}{20,105,176}
\colorlet{numb}{magenta!60!black}

\newtoggle{draft}
\settoggle{draft}{false}
\newtoggle{camera}
\settoggle{camera}{true}

\def\BibTeX{{\rm B\kern-.05em{\sc i\kern-.025em b}\kern-.08em
    T\kern-.1667em\lower.7ex\hbox{E}\kern-.125emX}}

\setlist[itemize]{leftmargin=*}
\setlist[enumerate]{leftmargin=*}

\begin{document}

\title{Trusted Container Extensions for Container-based Confidential Computing\\
}

\iftoggle{draft}{
    \newcommand{\delete}[1]{}
    \pagestyle{plain}
    \author{\IEEEauthorblockN{Anonymous}
    \IEEEauthorblockA{}
    }
}{
    \newcommand{\delete}[1]{}
}

\iftoggle{camera}
{\author{
\rm{Ferdinand Brasser, Patrick Jauernig, Frederik Pustelnik,}\\
\rm{Ahmad-Reza Sadeghi, Emmanuel Stapf}\\
\textit{Technical University of Darmstadt, Germany}\\
\normalsize{\{ferdinand.brasser, patrick.jauernig, emmanuel.stapf}\}\normalsize{@sanctuary.dev}\\
\normalsize{\{ahmad.sadeghi}\}\normalsize{@trust.tu-darmstadt.de}}
%%%%%%%%%%%%%%%%%%%%%%%%%%%%%%%%%%%%
}

%\iftoggle{camera}{
%    \author{\IEEEauthorblockN{1\textsuperscript{st} Given Name Surname}
%    \IEEEauthorblockA{\textit{dept. name of organization (of Aff.)} \\
%    \textit{name of organization (of Aff.)}\\
%    City, Country \\
%    email address or ORCID}
%    \and
%    \IEEEauthorblockN{2\textsuperscript{nd} Given Name Surname}
%    \IEEEauthorblockA{\textit{dept. name of organization (of Aff.)} \\
%    \textit{name of organization (of Aff.)}\\
%    City, Country \\
%    email address or ORCID}
%    \and
%    \IEEEauthorblockN{3\textsuperscript{rd} Given Name Surname}
%    \IEEEauthorblockA{\textit{dept. name of organization (of Aff.)} \\
%    \textit{name of organization (of Aff.)}\\
%    City, Country \\
%    email address or ORCID}
%    \and
%    \IEEEauthorblockN{4\textsuperscript{th} Given Name Surname}
%    \IEEEauthorblockA{\textit{dept. name of organization (of Aff.)} \\
%    \textit{name of organization (of Aff.)}\\
%    City, Country \\
%    email address or ORCID}
%    \and
%    \IEEEauthorblockN{5\textsuperscript{th} Given Name Surname}
%    \IEEEauthorblockA{\textit{dept. name of organization (of Aff.)} \\
%    \textit{name of organization (of Aff.)}\\
%    City, Country \\
%    email address or ORCID}
%    \and
%    \IEEEauthorblockN{6\textsuperscript{th} Given Name Surname}
%    \IEEEauthorblockA{\textit{dept. name of organization (of Aff.)} \\
%    \textit{name of organization (of Aff.)}\\
%    City, Country \\
%    email address or ORCID}
%    }
%}{}

\maketitle

\begin{abstract}
\input{sections/abstract}

\end{abstract}


\input{sections/intro}
\input{sections/bg}
\input{sections/adversary}

\input{sections/req}
\input{sections/design}
\input{sections/implementation}
\input{sections/evaluation}
\input{sections/related_work}
\input{sections/conclusion}

\bibliographystyle{abbrv}
%\printbibliography
\bibliography{library_compact}

\end{document}

%% file: sections/abstract.tex
Cloud computing has emerged as a corner stone of today's computing landscape. More and more customers who outsource their infrastructure benefit from the manageability, scalability and cost saving that come with cloud computing. Those benefits get amplified by the trend towards microservices. Instead of renting and maintaining full VMs, customers increasingly leverage container technologies, which come with a much more lightweight resource footprint while also removing the need to emulate complete systems and their devices. 

However, privacy concerns hamper many customers from moving to the cloud and leveraging its benefits. Furthermore, regulatory requirements prevent the adaption of cloud computing in many industries, such as health care or finance. Standard software isolation mechanisms have been proven to be insufficient if the host system is not fully trusted, e.g., when the cloud infrastructure gets compromised by malicious third-party actors. Consequently, confidential computing is gaining increasing relevance in the cloud computing field. 

We present Trusted Container Extensions (\arch), a novel container security architecture, which combines the manageability and agility of standard containers with the strong protection guarantees of hardware-enforced Trusted Execution Environments (TEEs) to enable confidential computing for container workloads. \arch provides significant performance advantages compared to existing approaches while protecting container workloads and the data processed by them. Our implementation, based on AMD Secure Encrypted Virtualization (SEV), ensures integrity and confidentiality of data and services during deployment, and allows secure interaction between protected containers as well as to external entities. Our evaluation shows that our implementation induces a low performance overhead of 5.77\% on the standard SPEC2017 benchmark suite.

%We also evaluate network throughput, where we measure an average overhead of 22.1\% for NGINX and 13.36\% for Apache.

%During the recent years, the growth of cloud computing lead to more and more customers outsourcing their infrastructure. One way for customers to package their workloads is to use containers instead of virtual machines because of their lightweight resource footprint. Providing multiple userland instances removes the need to emulate complete systems and their devces. However, working on sensitive data in the cloud leads to a risk of exposing data to others. Software isolation means prove to be infeasible if the host system is not trusted. A threat emerges if cloud providers might try to extract information about user workloads either by intend or if they have been infiltrated by third-parties.

%We present a novel container security architecture, which protects user's containers from maliciously acting cloud providers by leveraging AMD Secure Encrypted Virtualization (SEV). Current confidential computing solutions often have to take a trade-off between security and performance, while SEV is able to achieve both. Our implementation keeps container data safe at all times, and allows the user to securely communicate with his remote workload. Our evaluation shows that our implementation induces an performance overhead of 5.77\% on SPEC2017. We also evaluate network throughput, where we measure an average overhead of 22.1\% for NGINX and 13.36\% for Apache.

%% file: sections/intro.tex
\section{Introduction}
For over a decade, there is a continuous trend towards cloud computing, which allows customers to leverage capability and cost advantages. Cloud computing evolved with the advent of virtualization~\cite{xen}. Virtual machines (VMs) enabled Infrastructure-as-a-Service (IaaS) which allows businesses and users to outsource pre-existing workloads to the cloud. However, in recent years, the trend in cloud computing has shifted from VM-based offerings to more lightweight solutions, in particular, container technologies~\cite{docker,kubernetes,googlecontainer,amazoncontainer}. 

Containers, such as Docker~\cite{docker}, provide multiple separated user-space instances, which are isolated from each other and the host system through kernel software mechanisms. By running directly on the host system, containers do not need complex device emulation, large virtual machine disk files and packages pre-configured applications with all their dependencies which makes them an attractive choice for fast deployment of webservices. 
%Yet, containers enable various applications to execute independently on a shared host system. 
\delete{Moreover, another advantage of containers is their ability to package pre-configured applications with all libraries they depend on into bundles to create largely self-contained (micro-) services, which are easy to deploy. Orchestration systems, such as Kubernetes~\cite{kubernetes}, provide manageability and agility for these (micro-) services.}
Cloud providers today recognized this trend and offer customers the possibility to deploy and manage containers in the cloud~\cite{googlecontainer, amazoncontainer}, known as called Container-as-a-Service (CaaS), with Docker being currently the most popular container ecosystem~\cite{dockermarketshare, docker}.

Despite offering many advantages, using cloud services introduces a risk of data being exposed to third parties or services being compromised~\cite{data-breaches}.
Furthermore, regulatory policies~\cite{duncan2018can} restrict the adoption of cloud services for many industries, such as health care or finance. Even if the cloud service provider (CSP) is considered trustworthy, the CSP’s infrastructure might be compromised, e.g., by insiders such as maliciously acting administrators and employees, nation state actors demanding access by law, as well as third-party entities. 
\delete{Exploiting software vulnerabilities allows an adversary to compromise co-tenants’ applications directly or gain control over the host system’s privileged software in order to compromise the integrity and confidentiality of hosted workloads.} 
While the hypervisor software components, which are used to control and manage VMs, have been subject to various attacks~\footnote{\tiny CVE-2017-10912, CVE-2017-10918, CVE-2017-10920, CVE-2017-10921}, the attack surface in CaaS settings is even larger as a typically large and complex operating system kernel is responsible for managing and isolating the containers~\footnote{\tiny CVE-2015-8967, CVE-2016-10229, CVE-2016-7117, CVE-2017-0335, CVE-2017-0427, CVE-2017-0561}.

In recent years, confidential computing has gained relevance in the realm of cloud computing in a pursue to enable the trustworthy outsourcing of sensitive data and services to the cloud, while eliminating the requirement to trust the CSP. Leveraging hardware-enforced Trusted Execution Environments (TEEs), the user's workloads are protected inside isolated compartments, called \emph{enclaves}, which are secure even if the host’s privileged software is compromised or controlled by a malicious entity. 
Various TEE architectures have been proposed by academic research~\cite{sanctuary, cure, sanctum, keystone, aegis, flicker, bastion, iso-x, trustice}, while commercially available and widely deployed TEEs are Arm TrustZone~\cite{arm-trustzone}, Intel SGX~\cite{sgx} and AMD SEV~\cite{sev}. Just Recently, Intel and Arm announced new TEE architectures named Intel Trust Domain Extensions (TDX)~\cite{tdx} and Arm Confidential Compute Architecture (CCA)~\cite{cca}. 
\delete{They are all targeting specific use cases and have individual shortcomings. Arm TrustZone originates from a mobile setting and offers only a single TEE, and it relies on a trusted operating system to isolate application and services inside its single TEE. Intel SGX is designed to isolate parts of an application in the user space. AMD SEV, in contrast, is designed to isolate entire VMs from an untrusted hypervisor aiming to provide a solution for IaaS cloud settings. TEE architectures providing similar capabilities as AMD SEV have been announced by Intel, called Trust Domain Extensions (TDX)~\cite{tdx}, as well as Arm, called Confidential Compute Architecture (CCA)~\cite{cca}.}
However, none of the available TEE architectures is designed to isolate container workloads and to securely orchestrate and manage those. The demand for securely isolated container is shown by efforts to isolate containers in enclaves using Intel SGX~\cite{arnautov2016scone} or Arm TrustZone~\cite{tz-container}, however, these approaches either suffer from unpractical performance overheads or cannot protect from malicious cloud providers.
%The need for a practical and secure isolation solution for containers is shown by Google’s efforts to combine its CaaS with its confidential computing services~\cite{cc-gke}. Unfortunately, these developments are proprietary to Google’s cloud. %and no details about this approach are publicly available.

In this paper, we present Trusted Container Extensions (\arch), a novel security architecture providing strongly isolated containers that can be securely deployed and managed in the cloud. We leverage existing TEE architectures, such as AMD SEV, Intel TDX or Arm CCA, to ensure the integrity and confidentiality of applications and data in use and at rest. 
%By leveraging TEE architectures that can isolate entire VMs, such as AMD SEV, Intel TDX or Arm CCA, 
We protect containers in special-build lightweight VMs, called Secure Container VMs (\vms). \arch preserves the agility and manageability of containers by offering secure services for standard Docker containers. Using a single trusted VM per host system, \arch provides advanced security services to all \vms, including secure deployment, secure remote access, secure storage and secure communication between \vms. 

\noindent\textbf{Contributions.} The main contributions of our work are:
\begin{itemize}
    \item We present \arch, a novel security architecture for secure containers in the cloud. \arch provides integrity and confidentiality guarantees for containers executed in untrusted clouds at all times.
    \item Our implementation of the \arch architecture provides seamless integration into Docker, based on AMD SEV and the Kata Containers project. 
    %We extend Kata Containers in order to isolate containers in SEV VMs and provide protection for \emph{all} components. 
    %To the best of our knowledge, \arch is the first comprehensive SEV-based container architecture that can protect from untrusted hosts. 
    \item \arch provides a secure and transparent communication channel for secure containers, i.e., Docker cannot distinguish between locally or remotely executed containers.
    \item We thoroughly evaluate our implementation with respect to security and performance aspects. We analyze relevant attack vectors and explain how \arch protects against these threats. In our performance evaluation, which shows the practicability of our implementation, we evaluate computational-intensive workloads (SPEC2017 benchmark suite), network-intensive workloads (NGINX and Apache webserver) and memory-intensive workloads (Redis in-memory database). %\arch achieves an overhead of 5.77\% in SPEC2017, 22.1\% for NGINX request throughput and 13.36\% in Apache request throughput.
\end{itemize}

\delete{\textbf{Outline.} The remainder of this paper is organized as follows: First, in \Cref{background}, we provide an overview of relevant background on the technologies \arch builds upon. \Cref{adversary} presents our adversary model, whereby \Cref{requirements} lists all requirements \arch needs to fulfill. \Cref{design} details on \arch{}’s design. Our implementation is described in \Cref{implementation}. Afterwards, we analyze \arch{}’s security aspects discussing possible attacks and explain how these attacks are mitigated by \arch in \Cref{evaluation}. Next, the performance characteristics of \arch are evaluated (\Cref{evaluation}). We discuss potential extensions of \arch in \Cref{discussion} and give an overview of related work in \Cref{related}. Finally, we conclude in \Cref{conclusion}.}

%% file: sections/bg.tex
\section{Background}
\label{background}
In this section, we introduce the key technologies \arch is based on, which are: Trusted Execution Environments (TEEs), the security architecture AMD Secure Encrypted Virtualization (SEV) and Kata Containers.
%In this section, we give an overview of all background needed in order to understand the rest of this paper.

\subsection{Trusted Execution Environments}
Trusted Execution Environments (TEEs)~\cite{sgx, arm-trustzone, sev, sanctum, sanctuary, keystone, cure} are a type of security architecture which became prominent in recent years. TEEs securely isolate workloads from their underlying host system, such that they cannot be manipulated during run time. TEE implementations provide one or more secure execution environments, often called \textit{enclaves}. Enclaves run in parallel to the commodity operating system and applications, which are referred to as the Rich Execution Environment (REE). The Trusted Computing Base (TCB) includes all software and hardware which enforces the security guarantees of the TEE and which needs to be trusted inherently.
%are needed in order to provide all security guarantees by the TEE. 
TEEs try to achieve the following goals:
\begin{itemize}
\item \textbf{Strong Run-time Protection.}  Isolate enclaves from the REE at rest and at run time. Execution primitives, e.g., the register state or the memory of the TEE, must be isolated from the REE at all points in time.
%memory contents, the register state and caches should be isolated from the REE at any point in time. \\
\item \textbf{Verifiable State.} The TEE's boot process and state during must be externally verifiable at run time using \textit{attestation}.
%Often, it is desired to prove the secure state of a TEE to a third-party. The process of proving that an enclave was booted in a secure state is called \textit{attestation}. \\
\item \textbf{Small TCB.} The TCB must be kept minimal to reduce the risk of vulnerabilities that can lead to a TEE compromise.

%since a large codebase inevitably increases the risk of software vulnerabilities which might be exploited to compromise the TEE.
%Modern software has an ever-growing codebase and therefore a high risk for introducing errors and vulnerabilities. As it is favorable for a TEE to have an attack surface as small as possible, the TCB of the TEE should also be kept as small as possible. \\
\item \textbf{Backwards Compatibility.} A TEE should integrate into the existing hardware and software ecosystem as easily as possible in order to promote its adoption in practice.
%to attract developers. \\
\item \textbf{Low Performance Overhead.} Another important factor for the acceptance of a TEE in practice is that it only introduces a low performance overhead.
%In order to raise acceptance for developers, it is favorable for a TEE to introduce only a low performance overhead. \\
\end{itemize}

\subsection{AMD Secure Encrypted Virtualization}
AMD Secure Encrypted Virtualization (SEV)~\cite{sev, sev-es, snp} is a TEE architecture which targets cloud servers. SEV allows to protect multiple Virtual Machines (VMs) from a malicious host system or underlying hypervisor. SEV was built by reusing and extending existing features of AMD systems, namely, the Secure Virtual Machine (SVM) extension and the Secure Memory Encryption (SME)~\cite{sev} extension. Moreover, a co-processor is added to the System-on-Chip (SoC), called the Platform Security Processor (PSP)%, which is an Arm Cortex-A5 processor separated from the main CPU. 
In the following, we describe how these technologies are combined in SEV to protect sensitive VMs, even in the advent of strong adversaries or a malicious cloud provider.
%Protection of the VMs is achieved through two mechanisms: Memory encryption during runtime and attestation, which proves that a VM has been booted in an expected and secure state. Both of these features are handled by the Platform Security Processor (PSP) which represents SEV's TCB. 

\delete{\textbf{AMD Virtualization.} 
Secure Virtual Machine (SVM) is AMD's instruction set extension to provide hardware support for virtual machines. One of the key instructions newly introduced with SVM is the \texttt{vmrun} instruction which takes a Virtual Machine Control Block (VMCB) as an argument. When executing \texttt{vmrun}, the CPU transitions from \textit{host} mode to \textit{guest} mode and stores the register state of the host at a pre-defined location. Then, the guest register state is loaded from the VMCB. Besides the guest register state, the VMCB also contains an Address Space Identifier (ASID), which is unique to every VM and used to tag entries of a VM in the Translation Lookaside Buffer (TLB). Furthermore, the VMCB contains a pointer to the nested page tables, which map perceived guest physical memory to host physical memory. %As a result, every guest memory access requires two page table walks to become fully translated.
}

% \begin{figure}[h]
%     \centering
%     \includegraphics[width=0.35\textwidth]{images/sev.pdf}
%     \caption{AMD Secure Encryption Virtualization encrypts the VM data transparently by assigning VM Encryption Keys (VEKs) to Address Space Identifiers (ASIDs). }
%     \label{fig:sev}
% \end{figure}

\textbf{Memory Encryption.} 
Secure Memory Encryption (SME)~\cite{sev} encrypts the complete system memory in order to prevent live RAM introspection or cold-boot attacks~\cite{cold_boot}. In SEV, the SME feature is extended to provide memory encryption for sensitive VMs, whereby the memory of each VM is encrypted with a different key in order to achieve an isolation between the VMs and also the REE.
It is the VM's responsibility to define which pages are encrypted and which are shared with the hypervisor.
%, the boot code and the kernel need explicit support for SEV. 
%This is done with the C-bit (bit 47) in the page table entries of the VM. 
%If the C-bit is set during a memory access, the access will decrypt the memory content accordingly. 
The PSP exposes various commands to the hypervisor in order to set up SEV for a VM and to encrypt its initial boot code. The PSP generates and assigns each VM a unique VM Encryption Key (VEK), which is not accessible by any software running on the main CPU, and configures the memory controller accordingly. The binding between VM and VEK is done via the ASID which is used to index into the list of all currently usable VM encryption keys. An AES engine resides inside the memory controller, which encrypts data with 128 bit keys in Electronic Codebook (ECB) mode. %We illustrate the assignment of VEKs to VMs in~\Cref{fig:sev}.

\textbf{Attestation.} The attestation functionality for SEV VMs is provided by the PSP which is the first component to boot on the SoC. Every AMD SoC exports various public certificates, including a public Diffie-Hellman share (PDH) and a Chip Endorsement Key (CEK) for unique identification. Before launching a VM, the VM owner can request a signed attestation report of the PSP for the initial encrypted bootcode. If the verification of the attestation is successful, the owner can use the PDH to create a shared secret symmetric transportation key with the PSP, which is used to inject an encrypted secret into the VM. By leveraging the Diffie-Hellman scheme, only the target PSP is able to decrypt the secret and to inject the secret into the VM by encrypting it with the VEK.

\delete{\textbf{SEV Versions.} The initial version of SEV~\cite{sev}, introduced in 2016, lacked some necessary features for proper VM isolation. Memory encryption and attestation were already supported, however, the first version of SEV left the register state visible to the host system via the VMCB, leading to attacks~\cite{sev_attack_vmcb}.
%Consequently, attacks based on the register state emerged, which allowed to either run code execution or memory leakage attacks~\cite{sev_attack_vmcb}. Based on the accessible register state, an adversary could easily find gadgets within the VM's memory.
%, which are essentially a memory read into a register.
%Based on these gadgets and by manipulating the registers in the VMCB, a complete leakage of VM memory is possible. 
AMD countered these attacks with the next SEV version, Secure Encryption Virtualization - Encrypted State (SEV-ES)~\cite{sev-es}. SEV-ES encrypts and authenticates the register state of the VMCB during context switches in order to mitigate previously published attacks. However, SEV-ES only provided encryption but no integrity protection for the VM memory. Academia showed that the swapping of page-table entries and rollback attacks are possible because of SEV's missing integrity protection~\cite{sev_attack_integrity1, sev_attack_integrity2}. Further research showed that rollback attacks can also be performed on the PSP's firmware~\cite{sev_attack_psp}. The newest SEV version introduced by AMD, Secure Nested Paging (SEV-SNP)~\cite{snp}, fixes these problems by restricting the hypervisor's access to nested page tables during executing of a VM. Moreover, the ciphertext of a page is made depended on its memory location and SEV-SNP enforces that for every memory address, a SEV VM always reads the same memory value which it wrote last to the address.}

\subsection{Kata Containers}

Kata Containers \cite{kata} is a project which aims to add another level of isolation to Docker containers by executing containers within VMs. The project considers an adversary who is able to break out of the software isolation environment offered by containerization, and has the capability to escalates his privileges. The underlying hypervisor, in contrast to SEV, is assumed to be trusted. By encapsulating containers in VMs, the adversary only gains access to the information of the container VM. The host system and all other container VMs remain protected.
%The adversary only has access to a VM with non-confidential information, which leaves the host system safe in case of a kernel privilege escalation attack. 
Kata Containers integrate seamlessly into the existing Docker ecosystem. Kata consists of the kata-runtime, Kata VMs, and the kata-agent.

The \textbf{kata-runtime} process runs on the underlying host system and receives Open Catalog Interface (OCI) compatible commands from containers (a standardized high-level container runtime), which are translated into internal commands. The kata-runtime is also responsible for setting up and starting the hypervisor and all Kata VMs, including their devices.

The \textbf{Kata VM} is the VM in which a container is executed. It is based on a minimal Linux image, which offers as few services as possible. Also, the kernel is compiled to only support absolutely necessary drivers, most of them virtio drivers, which represent virtualization aware devices for better performance. Kata can be configured to use one of multiple hypervisor technologies, e.g. KVM~\cite{kvm}.

Finally, the \textbf{kata-agent} process runs within the Kata VM and receives commands from the kata-runtime. The kata-agent is based on runc, the default container runtime used by Docker. As a container runtime, it is responsible for isolating the container process and setting up all needed mechanisms, such as, namespaces, chroot, or seccomp.

%% file: sections/adversary.tex
\section{Adversary Model}
\label{adversary}
We assume \arch to be implemented on off-the-shelf cloud-targeted TEEs which provide strong protection for VM workloads. Currently, the only product-ready TEE which focuses on cloud servers is AMD SEV and thus, we inherit our adversary model from its newest version, SEV-SNP.\footnote{Recently new TEEs, Trust Domain Extensions (TDX)~\cite{tdx} and Arm's Confidential Compute Architecture (CCA)~\cite{cca} have been announced, which have high-level goals and adversary model similar to SEV.}

SEV assumes a strong adversary, which is able to fully compromise the system software including the kernel of the host system and even the hypervisor, thus, the adversary can read and even manipulate the complete system memory during runtime. Moreover, he is able to spawn malicious host processes, commodity VMs and even SEV VMs. Also, the adversary can monitor the complete network traffic, i.e., inject packets and impersonate the host to observe the traffic send by the SEV VMs. The adversary's goal is to infiltrate the SEV VMs or to manipulate the network traffic sent to SEV VMs in order to manipulate them or extract their sensitive data.
%Network traffic also contains confidential information, which might allow the adversary to take control over the SEV VMs
Furthermore, the adversary has physical access to the server, which allows him to perform non-invasive physical attacks such as bus snooping or cold-boot attacks~\cite{cold_boot}.

Aligned with the threat model of SEV and other industry TEEs, e.g., Intel SGX~\cite{sgx} or Arm TrustZone~\cite{arm-trustzone}, side-channel attacks are considered out of scope, including microarchitectural side-channel attacks (e.g., performed on the TLB~\cite{cache_attack_tlb} or cache~\cite{percival2005cache, brasser2017software, zhang2016truspy}), controlled side-channel attacks~\cite{controlled_sc, sev_attack_cntrl1, sev_attack_cntrl2}, and physical side-channel attacks~\cite{em_attack, mangard2008power}.\footnote{To protect against side channels orthogonal approaches, e.g., side-channel-resilient algorithms~\cite{primeprobe,constant-time-bernstein} or randomization~\cite{brasser2019dr,newcache}, were developed.} Further, we consider Denial-of-Service attacks out of scope, since an adversary with full control over a system can, for example, shut down the complete system.
%Furthermore, we also do not consider any fingerprinting attacks against SEV VMs. These attacks aim to detect which services are running inside the SEV VM. 
The PSP, the memory controller and its integrated AES engine represent the hardware TCB of \arch, whereas our newly introduced \textit{\vendorvm}, which we describe in more detail in~\Cref{design}, represents \arch's software TCB. Aligned with other TEEs~\cite{sgx, arm-trustzone, sev, sanctum, sanctuary, keystone, cure}, we assume this small set of hardware and software components to be functioning correctly and to be inherently trusted. %We assume that SEV is correctly initialized during boot.

%% not included
% fault injection attacks~\cite{biehl2000differential}
% SEV attacks:
%   I/O operations  

%The design of SEV states that the VEK is never accessible to x86 software. As previously mentioned, SEV also builds on software support. For this, we assume that the boot code and the kernel set up SEV correctly. All initial pagetables need to have the C-bit set for proper shielding from the hypervisor. This ensures that the hypervisor cannot tamper with the SEV VMs even during boot time.

%For our adversary model, we build upon the adversary model assumed by SEV and Kata Containers, and will also refine the models for our specific scenario. As for our adversary model, we assume full physical control over the host system. SEV assumes a very strong attacker, which can even read and even manipulate the complete system memory during runtime. Furthermore, we assume that the adversary can monitor the complete network traffic. He can also inject packets and impersonate hosts. The adversary's goal is to infiltrate the SEV VMs or manipulate network traffic, over which requests to the VMs are sent and received. Network traffic also contains confidential information, which might allow the adversary to take control over the SEV VMs. As we also assume the Kata Containers threat model, we also assume a malicious container, which aims to infiltrate the host. Her goal is to break out of the Kata Containers sandbox and to infiltrate other SEV VMs.

%% file: sections/req.tex
\section{Requirements Analysis}
\label{requirements}
In this section, we list all requirements a practical security architecture needs to fulfill, in order to providing strongly protected software containers. In the remainder of the paper, we will show how \arch meets these requirements.

\begin{enumerate}[label=R\arabic*]
    \item \label{req1} \textbf{Container Confidentiality:} Confidentiality of all data \& code inside containers must be ensured at all times.
    \item \label{req2} \textbf{Container Integrity:} Integrity of containers must be preserved at run time and at rest. Any tampering attempts should either be detected or prevented.
    \item \label{req3} \textbf{Protection against Malicious Host and VM:} Confidentiality and integrity must be provided in presence of a malicious host system as well as in presence of malicious commodity and SEV VMs (\Cref{adversary}).
    \item \label{req4} \textbf{Secure Communication:} Secure communication channels between a container and its provider must be provided to enable a secure container management.
    \item \label{req5} \textbf{Secure Deployment and Attestation:} All containers must be started in an expected state, and the container state must be remotely verifiable through attestation.
    \item \label{req6} \textbf{Flexibility and Usability:} The solution must be highly usable and integrate into the existing software ecosystem in order be easily adoptable in practice.
    \item \label{req7} \textbf{Off-the-shelf Hardware:} The solution should not require hardware modifications. Instead, only off-the-shelf hardware platforms should be used to also being able to upgrade already manufactured cloud servers.
    \item \label{req8} \textbf{Low Performance Overhead:} The solution should induce only a moderate performance impact to achieve a reasonable trade-off between security and performance.
\end{enumerate}

%\item \label{req5} \textbf{Secure Deployment and Attestation:} All containers must be started in an expected state. Moreover, the container state must be remotely verifiable through attestation.
%\item \label{req6} \textbf{Flexibility and Usability:} The architecture must be highly usable and integrate well into the existing ecosystem in order to achieve high adoption in practice.
%Users wish to protect their existing workloads without ease. Security solutions have to integrate well into the existing ecosystem.
%\item \label{req7} \textbf{Practicability:} It is favorable that our implementation can be used in the real world and builds upon available technology.
%\item \label{req8} \textbf{Low Performance Overhead:} The solution should induce only a moderate performance impact Our implementation has a direct focus on cloud computing. Also, cloud providers aim to use the full potential of their hardware. Therefore, users and cloud providers wish to achieve low performance impact.

%% file: sections/design.tex
\section{Design}
\label{design}
In this section, we describe the design of Trusted Container Extensions (\arch). We first give an overview of the system components in our architecture. Next, we describe the lifecycle of a container secured with \arch and the provided secure-channel service.

The goal of \arch, as depicted in~\Cref{fig:design}, is to securely deploy sensitive containers on untrusted machines in cloud environments which provide TEEs~\cite{snp, cca, tdx}. 
The key idea of \arch is to execute every sensitive container in a single protected VM provided by the underlying TEE. The secure container environments, which we call \textit{Secure Container VM (\vm)}, are protected from the host system, the underlying hypervisor and all other \vms. Every \vm is securely deployed by a trusted VM called the \textit{\vendorvm}, which is deployed and managed by a trusted third party.
%The \vendorvm is securely deployed by a in turn also needs to be securely deployed. This process is done by a trusted third-party system, which deploys these trusted VM on every container host system. 
In the following, we describe each system component in our architecture:

\begin{itemize}
    \item The \textbf{Host System} is the system which hosts the commodity (unprotected) VMs, Secure Container VMs and the \vendorvm. As it is part of a cloud service infrastructure, multiple Host System instances can exist. In our architecture, as described in~\Cref{adversary}, we assume that the Host System is untrusted, as it may act malicious intentionally or could be compromised by an attacker.
    \item  The \textbf{Secure Container VMs} host all sensitive containers. They utilize the protection capabilities of the underlying TEE to achieve a strong isolation from the regular Rich Execution Environment (REE) of the Host System. In \arch, the Host System is capable of hosting many \vms simultaneously.
    \item The \textbf{Root VM} securely deploys \vms on the system, verifies their boot process and offers additional security services, such as the establishment of secure communication channels between secured containers. On each Host System, one instance of the \vendorvm exists. The \vendorvm is the trust anchor of \vms, similar to the Quoting Enclave in Intel SGX.
    \item The \textbf{Container Owner} is the cloud tenant who wants to deploy its sensitives container to the Host System. After deployment, \arch enables the Container Owner to securely manage its containers and exchange sensitive data with it. 
    %Our architecture also allows for secure data exchange between the Container Owner and his deployed container instance.
    \item The \textbf{Deploy System} (operated by the trusted third party) acts as a root-of-trust, and is responsible for securely deploying the \vendorvm to every Host System.
    \item The \textbf{TEE Primitives} are an abstract representation of the combined hardware and software components which enable the TEE functionality on the platform, and which provide basic security functionalities, such as, creating attestation reports or injecting secrets into its enclaves.
\end{itemize}

\subsection{Secure Container VM Lifecycle}
The lifecycle of an \vm consists of 4 steps (\Cref{fig:design}).

\textbf{\vendorvm Deployment.} The first step, which is only done once for every Host System, is the secure deployment of the \vendorvm performed by the trusted third-party Deploy System \circled{1}. First, the Deploy Service requests an attestation report from the TEE Primitives of the Host System. After the attestation report has been validated, the Deploy Service creates an identity $Cert_{RootVM}$ in the form of a certificate and injects it in the \vendorvm via the TEE Primitives.
%SEV secret injection process.

% \begin{figure}
%     \centering
%     \includegraphics[width=0.3\textwidth]{images/design.pdf}
%     \caption{High-level design of \arch showing the main system components and lifecycle of a Secure Container VM.}%
%     \label{fig:design}%
% \end{figure}

\textbf{Secure Container VM Creation.} Before an \vm can be created, the Container Owner uploads a confidentiality- and integrity-protected container image to the Host System. In order to create an \vm, the Container Owner requests the \vendorvm to attest the newly created protected VM. Moreover, the Container Owner sends his certificate $Cert_{Owner}$ to the \vendorvm \circled{2}. As the Container Owner can also request the certificate $Cert_{RootVM}$ of the Root VM, she can create an encrypted and authenticated channel for a secure communication with the container. If the attestation of the newly created \vm was successful, the \vendorvm creates a certificate $Cert_{VM_{N}}$. This certificate acts as an identity for the \vm. Then, the \vendorvm injects the secrets $Cert_{RootVM}$, $Cert_{VM_{N}}$ and $Cert_{Owner}$ into the \vm \circled{3}. Hence, the Container Owner knows exactly which \vm was created for her and the \vm knows who its owner is.

\textbf{Secure Container VM Communication.} When the \vm has been booted, the Container Owner establishes a secure channel to the \vm \circled{4} which is mutually authenticated using the previously distributed certificates. The channel is used by the Container Owner to securely exchange commands and data with the \vm.

\textbf{Secure Container VM Execution.} After establishing a secure communication channel to the \vm, the Container Owner instructs the Host System to load the protected container image into the \vm. Using the secure channel, the Container Owner sends the image key $Key_{Image}$ to the VM, which is used to load the image. After the image has been loaded, the Container Owner instructs the \vm to finally execute the container image.

\subsection{Secure Channel Service}
Besides secure deployment, the \vendorvm offers run-time services for secure-channel establishment between containers. Channel establishment uses container-level mutual authentication. It is also possible to determine the owner of the containers respective \vm, such that a container can also decide to only handle a subset of requests from certain owners.

In order to establish a secure communication between the two containers $C_1$ and $C_2$, the following steps are needed.
%The following steps have to be executed if a Container VM $VM_1$ wants to connect to $VM_2$:
\begin{enumerate}
    \item The \vm uses $Cert_{RootVM}$ in order to establish a secure channel to the \vendorvm. Then, \vm registers with a publicly visible name at the \vendorvm.
    \item The \vm requests $Cert_{VM_{2}}$ \& $Cert_{Owner_{2}}$ from the \vendorvm.
    \item $C_1$ establishes a secure authenticated channel (with $Cert_{VM_{2}}$) to $C_2$.
    \item If $C_2$ only wants to handle certain requests for specific owners, it can request $Cert_{Owner_{1}}$ of $C_1$ from the \vendorvm. During the secure-channel establishment, it then checks which requests by this party are allowed.
\end{enumerate}

%% file: sections/implementation.tex
\section{Implementation}
\label{implementation}
\begin{figure*}
\begin{minipage}{.65\textwidth}
    \centering
    \includegraphics[width=1\textwidth]{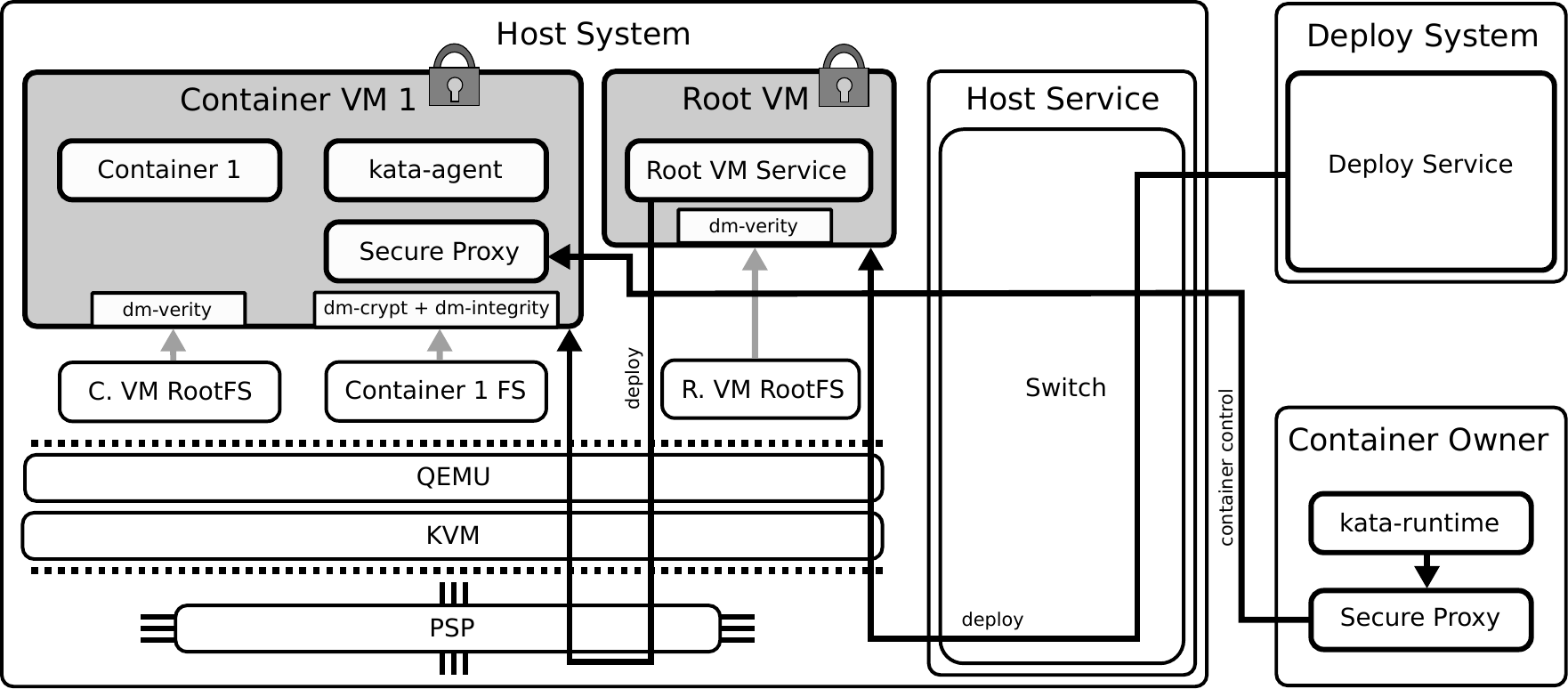}
    \caption{Implementation of \arch on an off-the-shelf AMD SEV host platform.}%
    \label{fig:implementation}%
\end{minipage}\hfill%
\begin{minipage}{0.27\textwidth}
    \centering
    \includegraphics[width=1\textwidth]{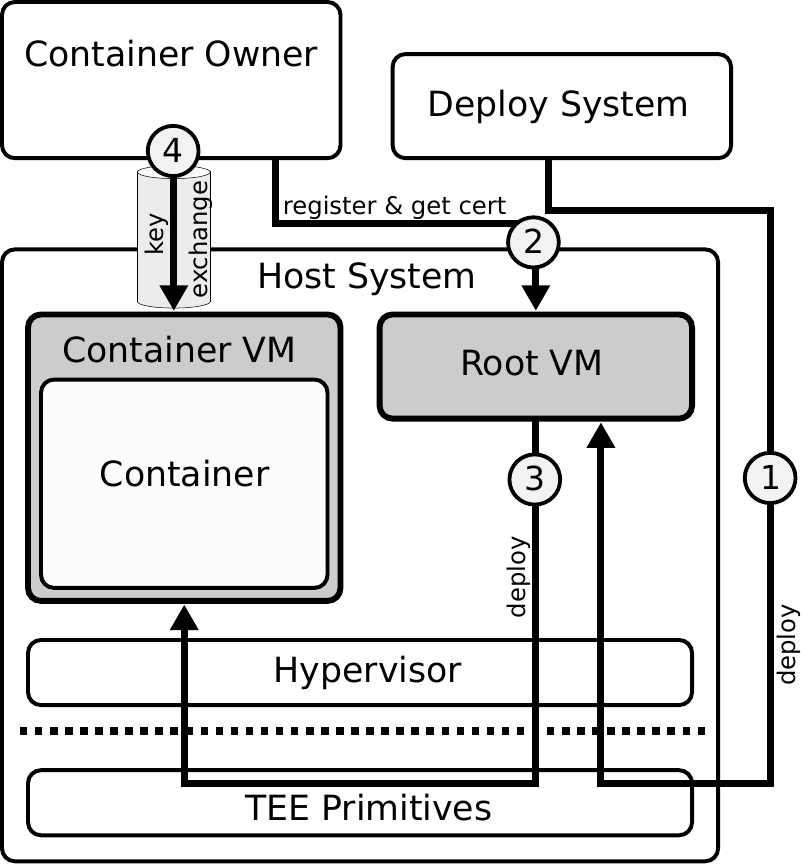}
    \caption{High-level design of \arch.}%
    \label{fig:design}%
\end{minipage}
\vspace{-0.4cm}
\end{figure*}
We implemented a prototype of \arch for Linux using AMD Secure Encrypted Virtualization (SEV), hence, fulfilling \ref{req7}. An overview of our prototype in shown Figure~\ref{fig:implementation}. We implemented container management using Kata Containers~\cite{kata} in combination with Docker~\cite{docker} as the container ecosystem. In the following, we first provide details on \arch's main system components and then, describe the \vm startup, the container images and \arch's role management.

\textbf{Container Owner.} The Container Owner system represents the cloud tenant that uses the provided cloud container service. We extended the existing Kata Containers architecture by executing the Kata VM on another system than the Container Owner's system. Running the kata-runtime on the Container Owner's system results in a completely transparent solution for Docker, such that Docker cannot differentiate between a locally running container and a remote container in the cloud. Docker only interacts with the kata-runtime via standardized Open Container Interface (OCI) commands and does not manage the execution of containers itself, but delegates this functionality to the runtime. This allows the Container Owner to manage the container state with existing tools, such that \ref{req6} is fulfilled. Normally, kata-runtime manages the hypervisor on which the \vm is running, as well as the kata-agent within the VM. Therefore, it is security-critical to verify authenticity of all requests sent to them and to secure the communication between them, otherwise these requests could be forged and the container could be compromised. We implement a secure proxy on both sides, such that all commands from the Container Owner are sent via an authenticated secure channel.

\textbf{Host System.} The Host System runs the Host Service, which manages the lifecycle of \vms and configures %QEMU as the Virtual Machine Monitor (VMM) and 
KVM as the hypervisor accordingly. Furthermore, the Host Service includes a switch, which routes incoming connections to a destination \vm. Container Ownerscan reach \vms and the \vendorvm via this exposed network interface. The Host Service is also responsible for container image management. It exposes a HTTPS server over which Container Owners can upload encrypted images. During \vm creation, the Host Service loads an image into a 9P filesystem, which is shared between host and VM.

\textbf{Deploy System.} The Deploy System hosts a Deploy Service, which builds upon sev-tool~\cite{sevtool}.%, a reference implementation by AMD for attesting SEV VMs.

\textbf{\vendorvm.} The \vendorvm%, which is based on Debian Buster in our prototype, 
is responsible for \vm attestation. Like the Deploy Service, it uses sev-tool for this process. The \vendorvm provides a secure-channel-creation service for establishing secure connections between containers. Our implementation is connection-oriented, analogous to existing network programming interfaces. We provide a simple Golang API for developers. 
%We show examples using our API in the appendix (\Cref{appendix}). 
The \vendorvm is conceptually similar to, e.g., Intel SGX's Quoting Enclave, and can be deployed at manufacturing time via board support packages (BSPs).
%or via drivers.

\textbf{\vm.} The kernel for the \vm is configured to only support the minimal set of devices needed for Kata Containers to function properly, i.e., the following set of QEMU devices: amd-iommu, pci-bridge, virtio-blk, virtio-scsi, virtio-9p, vhost-vsock-pci, virtconsiole, virtserialport, virtio-net.

\subsection{\vm Startup}
We implemented a Secure Boot process for every \vm, ensuring that every \vm is initially in a secure state. %In a Secure Boot process, each boot stage verifies the next stage, such that a chain-of-trust is ensured, where each stage is trusted because it has been verified by the previous stage. 
\arch uses PSP as the root for the secure boot chain-of-trust, which attests the UEFI boot code of each \vm.
%Our Secure Boot implementation uses the PSP as the root of this verification process, as it attests the UEFI boot code. 
All attestation reports are verified by sev-tool, an official implementation for SEV attestation by AMD. We implemented UEFI using Open Virtual Machine Firmware (OVMF).%, which is a port of Intel's tianocore firmware to QEMU. 
After OVMF has been booted, OVMF loads the Linux Kernel into encrypted memory and calculates its SHA256 hash. We modified OVMF, such that this hash is directly embedded in OVMF, and therefore, also verified by SEV attestation. Furthermore, we modified OVMF such that it can only load a kernel with an embedded hash. We also embed the kernel parameters within the image, as passing these from the host can lead to a potential compromise of the kernel. As the kernel will later pass execution to binaries of the \vm filesystem, we also have to protect the \vm filesystem from manipulation. We leverage  dm-verity for securing \vm images.%, as images only need to be read-only for our implementation. 
dm-verity is the de-facto standard for filesystem integrity checks on Linux, and is, e.g., used on Android systems~\cite{android_dmverity}. %dm-verity checks the filesystem integrity via a hash tree upon every filesystem read access. 
For ensuring integrity of the \vm's container images, we leverage dm-integrity in combination with dm-crypt, in order to form an authenticated encryption mechanism. We use AES with XTS as chaining mode for encryption, and HMAC-SHA256 for integrity tag computation.

\subsection{Container Images}
Docker splits container images into multiple layers, where each layer corresponds to a step in the image building process. Using this, Docker can store layers which are used by multiple images only once in order to save space. Images will then be reconstructed on-demand before the image is executed. Furthermore, Docker will add a writable top layer, which is then associated with a container instance. Since this is a key feature of Docker, our architecture also needs to implement this functionality. For this, we build upon the devicemapper functionality and its snapshot feature. Using this, we create a block overlay file in which all modifications of the encrypted container image will be stored in. Both files get loaded into the \vm upon start.

\subsection{Role Management}
In \arch, roles are represented by a Certification Authority hierarchy. The Root CA creates three Intermediate CAs, one for each type of entity, which sign TLS certificates for the according system types. The Intermediate \vendorvm CA also creates another CA, which is bound to a \vendorvm instance. This \vendorvm CA is responsible for creating \vm certificates, which are injected into the \vm and also sent to the Owner. Using this hierarchy, all systems can authenticate each other and identify the role of the opposite party. Systems can use mutual authentication where appropriate.

\noindent However, verifying the validity of \vendorvm certificates is challenging. As certificates should have a short validity timespan, distributing valid \vendorvm certificates, even if the \vendorvm itself has been shut down, leads to an unwanted bloat in the hierarchy tree. For this, revocation lists are the obvious choice, but they tend to grow rather large. Instead, the Deploy System provides a list of currently valid \vendorvm certificates, which can be pulled by the Container Owners' systems.

% \begin{figure}[h]
%     \centering
%     \includegraphics[width=0.3\textwidth]{images/ca.pdf}
%     \caption{Certification Authority hierarchy used for the secure role assignment implementation of \arch.}%
%     \label{fig:ca}%
%     \vspace{-0.3cm}
% \end{figure}

% \begin{lstlisting}[frame=single,caption={Secure Channel API for Go},breaklines=true,captionpos=b,label={channelapi},language=Golang,escapechar=|,style=mystyle]
% func Register(name string) error
% func GetRegisteredVMs() error
% func Dial(name string, owner *x509.Certificate) (net.Conn, error)
% func CheckForOwner(conn tls.Conn, name string) error
% func GetSelfOwner() (x509.Certificate, error)
% \end{lstlisting}
% 

%% file: sections/evaluation.tex
\section{Evaluation}
\label{evaluation}
In this section, we evaluate \arch with respect to security (\Cref{sec:evaluation:security}) as well as performance (\Cref{sec:evaluation:performance}). %First, we analyse the all security aspects, possible attack vectors and how we aim to secure containers against such kinds of attacks. Afterwards, we thoroughly evaluate the performance impact in computational and network throughput aspects.

\subsection{Security Analysis}
\label{sec:evaluation:security}
\arch provides bi-directional isolation: (1)~The containers are protected from accesses by external entities, including privileged entities. (2)~At the same time the host system, including the host's privileged software and other containers, are protected from unauthorized access by the container owner. Both scenarios are analyzed subsequently by distinguishing two types of adversaries.

The general concept and design of \arch can be instantiated with different TEE architectures, e.g., Intel TDX or Arm CCA. In this section we focus on the \arch implementation based on AMD SEV (\Cref{implementation}). However, the security arguments apply in similar ways for other TEE architectures.  

\textbf{Adversary Types.} Adversary Type~1~(A1) controls the host's system -- either a malicious cloud provider or an attacker that gained control over parts of the host system infrastructure -- and aims to gain access to a protected container. Adversary Type~2~(A2) is a malicious container owner that aims to break out of the virtualization-based sandbox.

\textbf{Container Isolation.} \arch leverages SEV to isolate containers. A1 would need to break the isolation guarantees of SEV, i.e., \arch's security is reduced to the security of the underlaying TEE architecture. Concretely, every \vm will be booted into a secure state. Furthermore, the memory controller ensures that every SEV VM uses memory which is encrypted through a different VEK. The VEK itself is never accessible to the hypervisor or other software, including other SEV VMs. OVMF and Linux are both SEV-aware, and they correctly set up all page tables in order to protect code and data from being shared with the hypervisor. The correct setup of an \vm is verified via attestation. Every SEV VM will be attested by another trusted entity. The \vendorvm is attested by the Deploy Service and all \vms on a Host System are attested by the \vendorvm, which fulfills \ref{req5}.

\textbf{TCB Size.} The TCB of \arch consists of the PSP and the Deploy System which is hosted by a trusted entity off platform. The \vendorvm which ensure the secure boot of all \vms is attested and securely booted by the Deploy System.
%Our changes to existing software, the Linux Kernel 5.11 and OVMF, are minimal with about 227 LOC in total. 
%The \vendorvm and Deploy Service share 1569 LOCs, which consists of code for attesting SEV VMs. 

%\noindent\begin{table}
%\centering
%\begin{tabular}{ c | c }
%\toprule
%Component & Added / Modified LoC \\
%\midrule
%OVMF & 115 \\ \hline
%Linux Kernel & 112 \\ \hline
%\vendorvm & 3767 \\ \hline
%Deploy Service & 2732 \\ \hline
%\bottomrule
%
%\end{tabular}
%\caption{Lines of Code required to implement our secure container architecture's trusted components.}
%\label{loc}
%\end{table}

\textbf{Secure Communication.} All network communication is secured via TLS. This ensures communication which is integrity-protected and confidentiality-protected from possible attacks of A1. Where required, we use mutual authentication. The Container Owner and his \vm use mutual authentication to be ensured that they communication with the real party. 

\textbf{Role Impersonation.} \arch encodes the roles of various system with a CA hierarchy. %Most promising targets for adversaries are the Deploy VM and \vendorvm, since both are responsible for SEV VM deployment. However, our implementation 
\arch checks that \vendorvm has a valid certificate which was signed by the \vendorvm Intermediate CA. This way, a malicious Host System A1 with a valid certificate in our CA cannot impersonate a \vendorvm. The certificate of the Deploy Service is validated in the same way. These mechanisms in combination with all communication secured fulfill\ref{req4}.

\textbf{Secure Storage.} As the container images need to be uploaded to the Host System prior to execution, it is necessary to store them securely, as they might contain sensitive information. For this, we leverage cryptsetup and integritysetup in order to create an authenticated container disk file. Before the \vm executes the first executable of the container image, the encryption key is passed to the \vm. As a result, only the \vm and the Container Owner can decrypt the container disk file, fulfilling \ref{req1} and \ref{req2}.

\textbf{Fake SEV.} An attack might want to fake the existence of SEV. An attacker A1 might emulate an SEV-enabled CPU, such that the SEV VMs assume to run within a genuine SEV VM. These kinds of attacks are prevented by SEV's design. Every AMD CPU has a unique certificate, which can be validated using AMD's servers. The encrypted SEV secret can only be decrypted by the PSP for which it was encrypted, as transport keys are used which in turn are encrypted with a secret key derived from the public DH share of the PSP. Only the target PSP knows the private part of the DH share and can successfully generate the secret to decrypt the transport keys.

\textbf{Malicious Container Owner.} Our implementation builds upon Kata Containers, which aims to isolate containers from the host system in case of a containerization escape through kernel exploits. Breaking out of the container software isolation environment does not offer a significant advantage to the adversary A2, as the \vm itself does not contain confidential information. The host memory, except shared MMIO, is not accessible to the SEV VM. Even if A2 is able to break out of the virtualization isolation (becoming A1), other \vms cannot be infiltrated, as we previously described. Hence, \ref{req3} is fulfilled.

\begin{table}
\centering
    \resizebox{\columnwidth}{!}{%
    \begin{tabular}{ c | c | c | c | c | c | c }
    \toprule
    vCPUs &  \shortstack{RAM\\ (GB)} & \shortstack{VM:\\ t./s} & \shortstack{+ \\(Base)} &  \shortstack{\vm:\\ t./s} & \shortstack{+ \\(Base)} & \shortstack{+ \\(VM)} \\
    \midrule
     
    1 & 2 &  10582 & 36.76\%  & 9783  & 41.58\% & 7.55\%  \\ \hline
    1 & 4 &  13362 & 20.15\%  & 12079 & 27.82\% & 9.6\%   \\ \hline
    2 & 2 &  13530 & 19.15\%  & 12433 & 25.7\%  & 8.11\%  \\ \hline
    2 & 4 &  14327 & 14.38\%  & 13251 & 20.81\% & 7.51\%  \\ \hline
    4 & 2 &  15413 & 7.89\%   & 13560 & 18.96\% & 12.02\% \\ \hline
    4 & 4 &  15819 & 5.47\%   & 13601 & 18.72\% & 14.02\% \\ \hline
    8 & 2 &  15744 & 5.91\%   & 13707 & 18.08\% & 12.93\% \\ \hline
    8 & 4 &  15619 & 6.66\%   & 13474 & 19.48\% & 13.73\% \\ \hline
    16 & 2 & 16283 & 2.7\%    & 12988 & 22.38\% & 20.23\% \\ \hline
    16 & 4 & 16363 & 2.21\%   & 13084 & 21.81\% & 20.0\%  \\ \hline
    \midrule
    
    \multicolumn{1}{c}{} & \multicolumn{1}{c}{} & \multicolumn{1}{c |}{}     & \multicolumn{1}{c |}{GM}  &  \multicolumn{1}{c |}{}   & \multicolumn{1}{c |}{GM} & \multicolumn{1}{c}{GM} \\ \hline
    \multicolumn{1}{c}{} & \multicolumn{1}{c}{} & \multicolumn{1}{c |}{}     & \multicolumn{1}{c |}{+8.82\%}   &  \multicolumn{1}{c |}{}   & \multicolumn{1}{c |}{+22.1\%}     & \multicolumn{1}{c}{+11.82\%} \\
    \bottomrule
    \end{tabular}%
    }
\caption{NGINX performance evaluation results for different resources. We compare throughput in \vms to Kata containers in non-SEV VMs. GM is the geometric mean.}
%The overall performance impact is measured with the geometric mean (GM).}
\label{nginxeval}
\vspace{-0.4cm}
\end{table}

\subsection{Performance Evaluation}
\label{sec:evaluation:performance}
We evaluated the performance of \arch on an SEV-capable system. We conducted experiments for computational performance, network throughput and database transaction performance, and show that our implementation satisfies \ref{req8}. As we improve over Kata containers, i.e., containers running within VMs for additional security, we compare against containers running in regular VMs without SEV.

\textbf{Experiment Setup.} For our experiments, we used a DELL PowerEdge R615 server with an AMD EPYC~7262 8-core CPU, $32\,GB$ memory and a $512\,GB$ SATA SSD. We connected the server via a $1\,GiB$ Ethernet to another system, which we ran the network throughput benchmarking tools, in order to eliminate any bottlenecks from running benchmarking tools on the same system. For this system, we used a Lenovo ThinkStation P330, equipped with an Intel i7-8700K 6-core CPU, 16GB memory and a $128\,GB$ SATA SSD.

\begin{figure*}
    \centering
    \footnotesize
    \subfloat[\centering NGINX 2GB VMs]{{
        \begin{tikzpicture} 
        \begin{axis} [legend columns=-1,
            scaled y ticks = false,
            every tick label/.append style={font=\footnotesize},
            xlabel={\# vCPUs},
            xlabel style={at={(0.5,0.15)}},
            ylabel={Requests/s},
            ylabel style={at={(0.0,0.5)}},
            xtick={1,2,4,8,16},
            xticklabels={1,2,4,8,16},
            ytick={9000, 11000, 13000, 15000, 17000},
            ymin=9000,
            ymax=18000,
            legend cell align=left,
            legend style={
                    at={(0.95,0.05)},
                    anchor=south east,
                    column sep=0ex,
                    nodes={scale=0.7, transform shape}
            },
            width=0.45*\textwidth,
            height = 3.2cm
            ]
            
        \addplot[smooth,mark=x,grey3] coordinates {
            (1,9783) (2,12433) (4,13560) (8,13707) (16,12988)
        }; \addlegendentry{\vm}
        
        \addplot[smooth,mark=x,black] coordinates {
            (1,10582) (2,13530) (4,15413) (8,15744) (16,16283) 
        }; \addlegendentry{non-SEV VM}
        
        \addplot[dashed, domain=1:16] {16734};
        \addlegendentry{Baseline}
        
        \end{axis}
        \end{tikzpicture}
    }}%
    \subfloat[\centering NGINX 4GB VMs]{{
        \begin{tikzpicture} 
        \begin{axis} [legend columns=-1,
            scaled y ticks = false,
            every tick label/.append style={font=\small},
            xlabel={\# vCPUs},
            xlabel style={at={(0.5,0.15)}},
            ylabel style={at={(0.15,0.5)}},
            xtick={1,2,4,8,16},
            xticklabels={1,2,4,8,16},
            ytick={9000, 11000, 13000, 15000, 17000},
            ymin=9000,
            ymax=18000,
            legend cell align=left,
            legend style={
                    at={(0.95,0.05)},
                    anchor=south east,
                    column sep=1ex,
                    nodes={scale=0.7, transform shape}
            },
            width=0.45*\textwidth,
            height = 3.2cm
            ]
            
        \addplot[smooth,mark=x,grey3] coordinates {
            (1,13362) (2,13251) (4,13601) (8,13474) (16,13084)
        }; \addlegendentry{\vm}
        
        \addplot[smooth,mark=x,black] coordinates {
            (1,12079) (2,14327) (4,15819) (8,15619) (16,16363) 
        }; \addlegendentry{non-SEV VM}
        
        \addplot[dashed, domain=1:16] {16734};
        \addlegendentry{Baseline}
        
        \end{axis}
        \end{tikzpicture}
    }}%
    %\vspace{0.2cm}
    
    \subfloat[\centering Apache 2GB VMs]{{
        \begin{tikzpicture} 
        \begin{axis} [
            every tick label/.append style={font=\small},
            xlabel={\# vCPUs},
            xlabel style={at={(0.5,0.15)}},
            ylabel={Requests/s},
            ylabel style={at={(0.0,0.5)}},
            xtick={1,2,4,8,16},
            xticklabels={1,2,4,8,16},
            ytick={4000, 5000, 6000, 7000, 8000, 9000},
            legend cell align=left,
            legend style={
                    at={(0.95,0.05)},
                    anchor=south east,
                    column sep=1ex,
                    nodes={scale=0.7, transform shape}
            },
            width=0.45*\textwidth,
            height = 3.2cm
            ]
            
        \addplot[smooth,mark=x,grey3] coordinates {
            (1,4725) (2,6645) (4,7680) (8,8630) (16,8387)
        }; \addlegendentry{\vm}
        
        \addplot[smooth,mark=x,black] coordinates {
            (1,6443) (2,7757) (4,8389) (8,8254) (16,8489) 
        }; \addlegendentry{non-SEV VM}
        
        \addplot[dashed, domain=1:16] {8542};
        \addlegendentry{Baseline}
        
        \end{axis}
        \end{tikzpicture}
    }}%
    \subfloat[\centering Apache 4GB VMs]{{
        \begin{tikzpicture} 
        \begin{axis} [
            every tick label/.append style={font=\small},
            xlabel={\# vCPUs},
            xlabel style={at={(0.5,0.15)}},
            xtick={1,2,4,8,16},
            xticklabels={1,2,4,8,16},
            ytick={4000, 5000, 6000, 7000, 8000, 9000},
            legend cell align=left,
            legend style={
                    at={(0.95,0.05)},
                    anchor=south east,
                    column sep=1ex,
                    nodes={scale=0.7, transform shape}
            },
            width=0.45*\textwidth,
            height = 3.2cm
            ]
            
        \addplot[smooth,mark=x,grey3] coordinates {
            (1,4806) (2,6875) (4,8564) (8,8325) (16,8391)
        }; \addlegendentry{\vm}
        
        \addplot[smooth,mark=x,black] coordinates {
            (1,6675) (2,8054) (4,7945) (8,8476) (16,8492) 
        }; \addlegendentry{non-SEV VM}
        
        \addplot[dashed, domain=1:16] {8542};
        \addlegendentry{Baseline}
        
        \end{axis}
        \end{tikzpicture}
    }}%
    \caption{NGINX and Apache performance evaluation for different resources. Both webservers are evaluated with different amounts of resources assigned to the \vms. The baseline is an unmodified container without a VM, non-SEV VM refers to a Kata container running in a regular VM.}%
    \label{fig:alleval}%
    \vspace{-0.3cm}
\end{figure*}
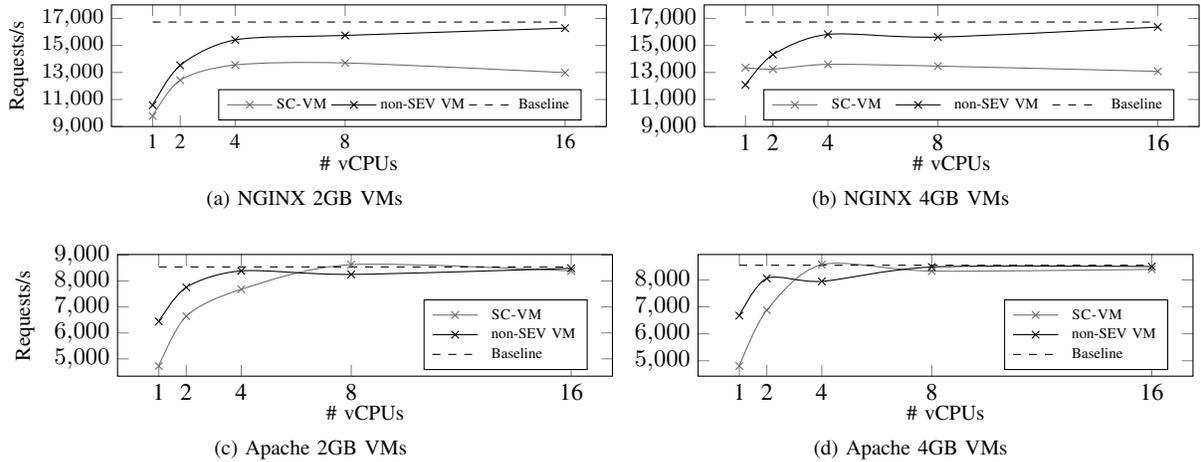

\textbf{NGINX.} As containers are often used to deploy web services, we first evaluate the currently most popular and widely distributed webserver NGINX. NGINX follows an asynchronous event-driven approach for connection handling and was explicitly developed for high performance. For throughput benchmarking, we use the Siege benchmark tool~\cite{siege}. For a more comprehensive view on how \vms perform, we evaluate these benchmarks with different amounts of resources assigned to the VMs. Table~\ref{nginxeval} and Figure~\ref{fig:alleval} show the resource-assignment and the results of our evaluation for NGINX. A standard VM induces an overhead of 8.82\% on NGINX request throughput, compared to the results of an unmodified container running NGINX on the host. A \vm induces an overhead of 22.1\% on throughput, which results in 11.82\% overhead relative to the standard VM. As NGINX frequently accesses memory, SEV's memory encryption is a likely cause for this overhead, especially due to the high number of requests/s that NGINX can manage.

\textbf{Apache.} We also evaluate the performance impact on Apache, another popular webserver. Apache forks its webserver process upon an incoming request, which leads to higher memory consumption than in event-driven model favored by NGINX. The evaluation results are shown in Table~\ref{apacheeval}. VMs induce an overhead of 7.21\% for Apache, while \vms induce an overhead of 13.36\%. Compared to the normal VM, the \vms induces an average overhead of 7.77\%. Interestingly, we observe performance gains for \vms with specific resource configuration in Figure~\ref{fig:alleval}, which we attribute to special beneficial scheduling and memory accesses conditions occurring when encapsulating Apache in a VM, outweighing the costs of memory encryption.

\begin{table}
\centering
    \resizebox{\columnwidth}{!}{%
    \begin{tabular}{ c | c | c | c | c | c | c }
    \toprule
    vCPUs &  \shortstack{RAM\\ (GB)} & \shortstack{VM:\\ t./s} & \shortstack{+ \\(Base)} &  \shortstack{\vm:\\ t./s} & \shortstack{+ \\(Base)} & \shortstack{+ \\(VM)} \\
    \midrule
     
    1 & 2 &   6443 & 24.57\% & 4725 & 44.68\% & 26.67\% \\ \hline
    1 & 4 &   6675 & 21.58\% & 4806 & 43.74\% &  28.0\% \\ \hline
    2 & 2 &   7757 &  9.19\% & 6645 & 22.20\% & 14.34\% \\ \hline
    2 & 4 &   8054 &  5.71\% & 6875 & 19.51\% & 14.64\% \\ \hline
    4 & 2 &   8389 &  1.79\% & 7680 & 10.09\% &  8.45\% \\ \hline
    4 & 5 &   7945 &  6.98\% & 8564 & -0.26\% & -7.79\% \\ \hline
    8 & 2 &   8254 &  3.37\% & 8630 & -1.06\% & -4.55\% \\ \hline
    8 & 4 &   8476 &  0.77\% & 8325 & 2.54\%  & 1.78\% \\ \hline
    16 & 2 &  8489 &  0.62\% & 8387 & 1.81\%  & 1.2\% \\ \hline
    16 & 4 &  8492 &  0.58\% & 8391 & 1,76\%  & 1.18\% \\ \hline
     \midrule
    
    \multicolumn{1}{c}{} & \multicolumn{1}{c}{} & \multicolumn{1}{c |}{}     & \multicolumn{1}{c |}{GM}  &  \multicolumn{1}{c |}{}   & \multicolumn{1}{c |}{GM} & \multicolumn{1}{c}{GM} \\ \hline
    \multicolumn{1}{c}{} & \multicolumn{1}{c}{} & \multicolumn{1}{c |}{}     & \multicolumn{1}{c |}{+7.21\%}   &  \multicolumn{1}{c |}{}   & \multicolumn{1}{c |}{+13.36\%}     & \multicolumn{1}{c}{+7.77\%} \\
    \bottomrule
    \end{tabular}%
    }
\caption{Apache performance evaluation results for different resources. We compare throughput in \vms to Kata containers on non-SEV VMs. GM is the geometric mean.}
\label{apacheeval}
\vspace{-0.4cm}
\end{table}

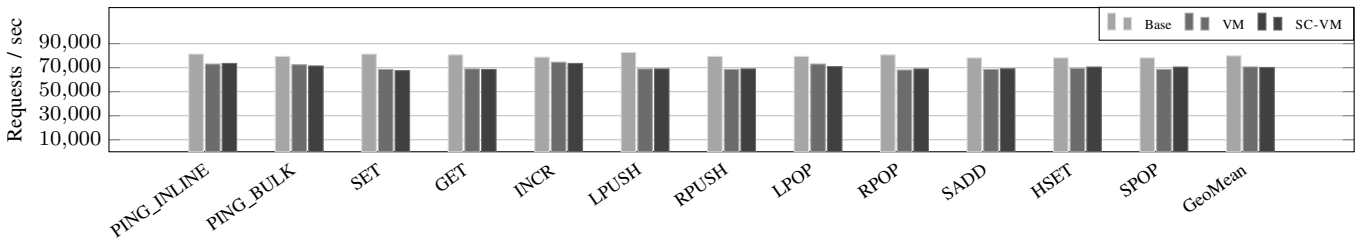
\begin{figure*}
    \centering
    \footnotesize
    \begin{tikzpicture}
        \begin{axis}[
            scaled y ticks = false,
            every tick label/.append style={font=\footnotesize},
            width  = \textwidth,
            height = 3.5cm,
            major x tick style = transparent,
            ybar=2*\pgflinewidth,
            bar width=5.5pt,
            ymajorgrids = true,
            ylabel = {Requests / sec},
            y label style={at={(0.0,0.5)}},
            symbolic x coords={PING\_INLINE,PING\_BULK,SET,GET,INCR,LPUSH,RPUSH,LPOP,RPOP,SADD,HSET,SPOP,GeoMean},
            x tick label style={font=\scriptsize,rotate=35,anchor=east,},
            xtick = data,
            ymin=0,
            ymax=120000,
            ytick={10000, 30000, 50000, 70000, 90000},
            legend cell align=left,
            legend style={
                    legend columns=-1,
                    at={(1,0.77)}, %at={(1,1.05)},
                    anchor=south east,
                    column sep=1ex,
                    nodes={scale=0.65, transform shape}
            }
        ]
            \addplot[style={grey1,fill=grey2,mark=none}]
                coordinates {(PING\_INLINE, 81300) (PING\_BULK, 79365) (SET, 81300) (GET, 80645) (INCR, 78740) (LPUSH, 82644) (RPUSH, 79365) (LPOP, 79365) (RPOP, 80645) (SADD, 78125) (HSET, 78125) (SPOP, 78125) (GeoMean, 79799)};
                
            \addplot[style={grey2,fill=grey3,mark=none}]
                coordinates {(PING\_INLINE, 72992) (PING\_BULK, 72463) (SET, 68493) (GET, 68965) (INCR, 74551) (LPUSH, 68965) (RPUSH, 68493) (LPOP, 72992) (RPOP, 68027) (SADD, 68493) (HSET, 69444) (SPOP, 68493) (GeoMean, 70623)};
                
            \addplot[style={grey3,fill=grey4,mark=none}]
                coordinates {(PING\_INLINE, 73529) (PING\_BULK, 71428) (SET, 67567) (GET, 68529) (INCR, 73529) (LPUSH, 68965) (RPUSH, 68965) (LPOP, 70921) (RPOP, 68942) (SADD, 69074) (HSET, 70529) (SPOP, 70529) (GeoMean, 70185)};
    
            \legend{Base, VM, \vm}
        \end{axis}
    \end{tikzpicture}
    \vspace{-0.3cm}
    \caption{Performance of redis-benchmark in a Secure Container VM (\vm) vs. Kata container vs. native container.}
    \label{fig:redisresults}
    \vspace{-0.3cm}
\end{figure*}

\textbf{Redis.} Redis is an in-memory key-value database service. In order to persistently store data, Redis forks itself and writes the state to storage in the background. The service supports a large variety of data structures, such as lists, hashtables and sets. %18 different commands are supported in total for manipulating these data structures. 
Redis offers a benchmarking tool, called redis-benchmark, which we used for benchmarking Redis within a \vm. Figure~\ref{fig:redisresults} shows the performance impact when running Redis within a normal non-SEV VM an within a \vm, both assigned with $4$ vCPUs and $8\,GB$ RAM. During experiments, it showed that Redis does not benefit from increased vCPUs and memory, in contrary to NGINX and Apache. Redis is highly dependent on single core performance. In our evaluation, we use $255$ concurrent connections and a default data size of $256\,bytes$. As our results show, a non-SEV VM introduces an overhead of 11.95\%, while a \vm introduces a overhead of 12.09\%.

\textbf{Use-Case: NGINX with CGI and MariaDB Database.} Web services often build upon multiple components in order to provide their functionality. For evaluating such a scenario, we built a web application hosted with NGINX, which is based on Python CGI. It takes a username and password as parameter and checks if both match the username and hash stored in a MariaDB database, which is running in a second \vm. During our experiment, we created two Container VMs with $4$ vCPUs and $4\,GB$ RAM. This configuration achieved 709 requests per second, while the baseline achieved 1174 requests per second. This results in a performance impact of 39.6\%, which can be partially attributed to the cost of having secure communication between the \vms.

\textbf{Secure Channel.} We also evaluated \arch's secure channel service for secure inter-container communication separately. For this, we developed a custom benchmarking tool, which repeatedly creates secure channel connection and sends a HTTP request. On average, a \vm was able to create 72.97 new connections per second. Further testing indicated that the bottleneck is during the data sending phase rather than the connection creation or TLS handshake phase. We attribute this behavior to our userland implementation. Userland implementations tend to have a drastic performance impact in contrast to virtualization-aware solutions. For example, we also evaluated network throughput for \vms with non-virtualization-aware devices our experiments showed that this setup only achieved around 15\% of the performance of the virtualization-aware setup. We believe that such a speedup is also possible for the secure channel.

\begin{figure*}
    \centering
    \footnotesize
    \begin{tikzpicture}[font=\footnotesize]
        \begin{axis}[
            every tick label/.append style={font=\footnotesize},
            width  = \textwidth,
            height = 3.5cm,
            major x tick style = transparent,
            ybar=2*\pgflinewidth,
            bar width=5.5pt,
            ymajorgrids = true,
            ylabel = {Run time [s]},
            ylabel style={at={(0.0,0.5)}},
            symbolic x coords={600.perlbench,602.gcc,605.mcf,620.omnetpp,623.xalancbmk,625.x264,631.deepsjeng,641.leela,648.exchange2,657.xz,GeoMean},
            x tick label style={font=\scriptsize,rotate=35,anchor=east,},
            xtick = data,
            scaled y ticks = false,
            ymin=0,
            ymax=1200,
            ytick={200, 400, 600, 800, 1000},
            legend cell align=left,
            legend style={
                    legend columns=-1,
                    at={(1,0.77)}, %at={(1,1.05)},
                    anchor=south east,
                    nodes={scale=0.65, transform shape}
            }
        ]
            \addplot[style={grey1,fill=grey1,mark=none}]
                coordinates {(600.perlbench, 364) (602.gcc,504) (605.mcf,679) (620.omnetpp,541) (623.xalancbmk,267) (625.x264,365) (631.deepsjeng,456) (641.leela,473) (648.exchange2,291) (657.xz,885) (GeoMean, 450)};
                
            \addplot[style={grey2,fill=grey2,mark=none}]
                coordinates {(600.perlbench, 367) (602.gcc,507) (605.mcf,680) (620.omnetpp,597) (623.xalancbmk,276) (625.x264,359) (631.deepsjeng,460) (641.leela,472) (648.exchange2,379) (657.xz,901) (GeoMean, 474)};
                
            \addplot[style={grey3,fill=grey3,mark=none}]
                coordinates {(600.perlbench, 375) (602.gcc,504) (605.mcf,698) (620.omnetpp,610) (623.xalancbmk,285) (625.x264,356) (631.deepsjeng,452) (641.leela,473) (648.exchange2,378) (657.xz,898) (GeoMean, 476)};
                
            \addplot[style={grey4,fill=grey4,mark=none}]
                coordinates {(600.perlbench, 371) (602.gcc,519) (605.mcf,691) (620.omnetpp,628) (623.xalancbmk,290) (625.x264,357) (631.deepsjeng,455) (641.leela,473) (648.exchange2,378) (657.xz,890) (GeoMean, 477)};
    
            \legend{Base, VM, SEV VM, \vm}
        \end{axis}
    \end{tikzpicture}
    \vspace{-0.3cm}
    \caption{SPEC2017 benchmark in a Secure Container VM (\vm) vs. Kata container on VM/SEV VMs vs. native container.}
    \label{fig:spec2017}
    \vspace{-0.3cm}
\end{figure*}
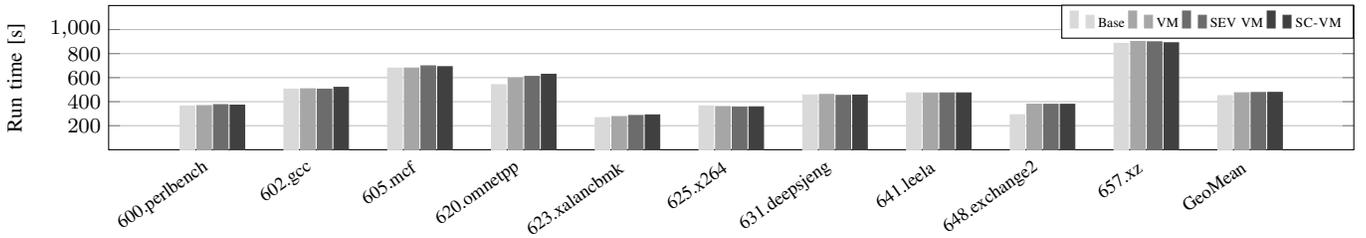

\textbf{SPEC2017.} We further evaluate \arch by measuring the computational performance impact introduced by SEV. For measuring computational overhead, we use SPEC2017, a commonly used benchmark tool. In order to thoroughly evaluate the overhead added by each isolation stage, we evaluate the overhead of virtualization, SEV and containerization within a SEV VM. We measured an average overhead of 4.92\% for a non-SEV VM. A SEV VM introduces an average overhead of 5.19\%.%, which is on par with other performance evaluations of SEV. 
The \vm introduces a slightly higher average overhead of 5.77\%. Hence, \arch's additional software isolation layer only introduces a very minor performance overhead.

\begin{table}[h]
\centering
\footnotesize
    \begin{tabular}{ l | c | c | c }
    \toprule
    Application & Disk Space & \shortstack{Idle \\ Memory Usage} & \shortstack{Load \\Memory Usage} \\
    \midrule
    NGINX (base) & 2.23MB & 2.2MB & 6.2MB \\ \hline
    NGINX in VM & 260MB & 758MB & 761MB \\ \hline
    \shortstack{NGINX in \vm} & 560MB & 1024MB & 1030MB \\
    \bottomrule
    \end{tabular}
\caption{Memory and disk space consumed by NGINX.}
\label{memusage}
\vspace{-0.3cm}
\end{table}

\subsection{Memory and Storage Overhead}
%We evaluated the memory and disk storage usage of our \arch implementation.
In Table~\ref{memusage}, we compare the disk usage and memory consumption during idle and load with a normal Ubuntu 20.04 VM. We further compare these results to the resource usage of \vms. \vms have a higher memory usage than a normal VM, since the usage of the devicemapper functionalities of the kernel requires additional memory. Our experiments showed that a \vm needs 1GB RAM in order to boot and to be able to load an encrypted container image.

%% file: sections/related_work.tex
\section{Related Work}
In this section, we compare \arch to related work, which we divide into the general approaches to protect legacy code using TEEs, and those works focusing on protecting containers using TEEs. Trusted Execution Environments (TEEs), in general, have been subject to extensive research in industry~\cite{sgx,arm-trustzone,sev}, as well as in the academic community~\cite{sanctuary, keystone, sanctum, cure, noorman2013sancus} and are therefore not discussed in this section. %In this work, we focus on actual ways to protect legacy code and containers, independent of the actual TEE, which have been primarily explored on commercial TEEs.

\label{related}
\subsection{Protecting Legacy Code}
A significant hurdle for adoption of confidential computing is the requirement to adapt code. While various examples of successful ports of complex services to TEEs exist~\cite{goltzsche2017trustjs,bayerl2020offline,wang2019running, baumann2015shielding}, it requires expert knowledge about TEEs. Hence, an ongoing trend is to enable unmodified software to run in a TEE. However, TEEs either only offer user-space execution, or require the deployment of an additional kernel and respective drivers by the developer. %In the following, we analyze approaches that aim to circumvent this problem.

%A first step into this direction was made by Microsoft Research, when they introduced Haven~\cite{baumann2015shielding}. Haven is a framework for shielding unmodified Windows applications from untrusted cloud providers using SGX. In order to support unmodified binaries, a complete Windows NT Kernel has been integrated within the SGX enclave to provide OS services. This NT kernel handles most syscalls, however, still needs to delegate a small set of syscalls to the host system. The Haven process in which the SGX enclave resides is a so-called Pico process, which allows a kernel module to implement custom syscall handlers for such processes. Pico processes were introduced by Microsoft for the first version of the Windows Subsystem on Linux, in order to offer custom syscall handler wrappers. Executing a complete NT kernel inside SGX also introduces a high performance impact, as SGX is not designed to run complete kernels and does not have syscall support within the enclave.

%In , researchers aimed to securely offer kernel services to enclaves, without the need to completely emulate or execute another kernel. 
Graphene-SGX~\cite{tsai2017graphene}, SGX-LKL~\cite{priebe2019sgx}, and Occlum~\cite{shen2020occlum} aim to securely offer kernel services to enclaves, without the need to completely emulate or execute another kernel. They leverage a so-called Library OS, which acts as a shielding layer for syscall services. Besides better performance, these approaches also offer a smaller TCB. For instance, Graphene-SGX implements most of kernel functionality in enclave code and syscalls are passed through to the host OS as needed. %In order to prevent Iago attacks from the host OS, a shielding layer performs sanity checks on return values of syscalls. Graphene-SGX also offers multi-process capabilities by starting multiple SGX enclaves with multiple Library OS instances, however, this approach showed to have weaknesses in performance.

In order to further secure SGX enclaves, PANOPLY~\cite{shinde2017panoply} aims to reduce the TCB by splitting enclaves into multiple micro containers, called \emph{microns}. PANOPLY also ensures secure inter-micron control-flow and data-flow and also provides a shielding layer for secure syscalls.

Nonetheless, all solutions building on Intel SGX suffer from the same problems. For one, Intel SGX requires code to be signed by Intel to load in release mode, preventing execution of user-supplied code. In order to circumvent this restriction, all above listed solutions mark a code region as writable and executable, and load the user-supplied code into this region. However, a restriction of SGX is that memory pages can set their attributes only once. This leaves all loaded usercode on pages with full permissions and introduces code injection attacks, which have been mitigated in regular software for over a decade. In addition, as all evaluations show, SGX-based solutions suffer from a significant performance impact.

Instead of relying on SGX, SEVGuard~\cite{palutke2019sevguard} isolates user-space applications by leveraging SEV. As the previously introduces approaches, SEVGuard also passes the syscalls to the host OS. However, SEVGuard does not suffer from the memory attribution problem, since SEV offers the capability to change memory attributes at any point in time. SEVGuard does not implement a secure storage mechanism or a shielding layer, which makes it vulnerable to Iago attacks.
Another vendor-independent virtualization approach to isolate applications is Sego~\cite{kwon2016sego}, which runs the OS and the program in different VMs. In order to handle syscalls within the user enclave, Sego also introduces a shielding layer.
In contrast, \arch is able to directly handle system calls within the enclave.

\subsection{TEE Container Runtimes}
SCONE~\cite{arnautov2016scone} and TZ-Container~\cite{tz-container} focus on securing containers from untrusted host systems. SCONE offers, like Graphene-SGX, a Library OS, which implements syscall functionalities and a shielding layer for host services. SCONE provides the integration of Docker containers in secure SGX enclaves. However, as other solutions, SCONE suffers from various problem introduced by Intel SGX, in particular, SGX enclaves have inflexible memory management (allocated at system boot) and developers need to explicitly implement SGX-specific multithreading by spawning threads in the host process that all enter the enclave. Further, due to SCONE's user-space threading model, SCONE cannot handle certain system calls such as \texttt{exec} or \texttt{fork}, hence, requiring modifications in many legacy applications.
In contrast, \arch leverages its own kernel in the Secure Container VM, handles system calls directly in the VM, and hence, offers near-native system call handling speeds. Also, \arch provides regular threading that does not require changes to the \vms, and provides highly-flexible resource allocation.
%, enabling allocating more memory if needed.

TZ-Container leverages management components in the TrustZone secure world to provide stronger isolation for containers and prevent page table modifications after configuration. For that, the normal-world kernel needs to be modified such that all page table modifications are trapped to the secure-world component. However, this directly implies that TZ-Container cannot protect from an untrusted cloud provider, as the kernel can be replaced with a non-trapping version. Contrary to TZ-Container, \arch also protects against malicious cloud providers and does not require kernel modifications.

%\textbf{Academic TEEs.} Academic research on TEEs focused on offering certain features and making TEE implementations more resilient against attacks.  \\

%% file: sections/conclusion.tex
\section{Conclusion}
\label{conclusion}
In this paper, we presented TCX, an architecture for secure deployment of containers on untrusted cloud systems. We leverage AMD SEV for our implementation, however, as our design is generic, it is also possible to support other TEE solutions if they meet the same requirements we outlined, e.g., the recently announced Armv9-A Confidential Computing Architecture (CCA), 
%using the so-called \emph{Realms}. Similarly, 
or Intel's recently announced Trust Domain Extensions (TDX).
%, an instruction set extension for secure virtual machines. 
Both of these architectures are the counterparts to SEV for Arm and Intel CPUs. Since they support the features needed for our architecture, it is possible to implement our architecture also on those security architectures. Finally, we showed that our architecture protects against many attack scenarios. We guarantee secure deployment of containers, their protection at runtime and to keep all data within the containers safe at all times. As our solutions builds on Kata Containers, it integrates seamlessly into the Docker architecture. Our evaluation shows a performance impact of 5.77\%, and a network throughput overhead of 22.1\% for NGINX and an overhead of 13.36\% for Apache.